\vspace{-20pt}
\documentclass[journal]{IEEEtran}
\usepackage{balance}
\vspace{-20pt}
\ifCLASSINFOpdf
\else
\usepackage[dvips]{graphicx}
\fi
\usepackage{subfigure}
\usepackage{cite}
\usepackage{mathrsfs}
\usepackage{amsmath,amssymb,amsfonts}
\usepackage{url}
\usepackage{algorithmic}
\usepackage{lipsum}
\usepackage{multicol}
\usepackage{graphicx}
\usepackage{epstopdf}
\usepackage{textcomp}
\usepackage{caption}
\usepackage{graphicx}
\usepackage{algorithmic}
\usepackage{algorithm}
\usepackage{bm}
\usepackage{array}
\usepackage{setspace}
\usepackage{makecell,multirow,diagbox}
\usepackage{textcomp,booktabs}
\usepackage[usenames,dvipsnames]{color}
\usepackage{colortbl}
\definecolor{mygray}{gray}{.9}
\definecolor{mypink}{rgb}{.99,.91,.95}
\definecolor{mycyan}{cmyk}{.3,0,0,0}

\newtheorem{lemma}{\textbf{Lemma}}

\ifCLASSOPTIONcompsoc
\usepackage[caption=true, font=footnotesize, labelfont=sf, textfont=sf]{subfig}
\else
\usepackage[caption=true, font=footnotesize]{subfig}
\begin{document}
	
	\title{\vspace{-0.5em}\LARGE Energy Efficient Design in IRS-Assisted UAV Data Collection System under Malicious Jamming }
	
	\author{Zhi Ji,
		Jia Tu, Xinrong Guan, Wendong Yang,	
         Weiwei Yang, and Qingqing Wu  \vspace{-2em}
     \thanks{
%The work of X. Guan was supported by the National Natural Science Foundation of China under Grant (62171461) and the Natural Science Foundation on Frontier Leading Technology Basic Research Project
%     	of Jiangsu under Grant (BK20212001). The work of Q. Wu was supported by the FDCT under Grant (0119/2020/A3, SKL-IOTSC(UM)-2021-2023), and the GDST under Grant (2021A1515011900, 2020B1212030003). (\textit{Corresponding author: Xinrong Guan}.)
     	
     	Zhi Ji, Xinrong Guan, Wendong Yang,  and Weiwei Yang are with the College of Communications Engineering, Army Engineering University of PLA, Nanjing, 210007, China (e-mail: jz20211009@163.com;  guanxr@aliyun.com;
     	ywd1110@163.com; wwyang1981@163.com). (\textit{Corresponding author: Xinrong Guan}.)
     
        J. Tu is with International Studies College, National University of Defense Technology, Nanjing, 210007, China (Email: tujia$\_$666@163.com).
        
        Q. Wu is with the State Key Laboratory of Internet of Things for Smart City, University of Macau, Macau, 999078, China (email: qingqingwu@um.edu.mo).
}
}

\maketitle
	
\begin{abstract}
In this paper, we study an unmanned aerial vehicle (UAV) enabled data collection system, where an intelligent reflecting surface (IRS) is deployed to assist in the communication from a cluster of Internet-of-Things (IoT) devices to a UAV in the presence of a jammer. We aim to improve the energy efficiency (EE) via the joint design of UAV trajectory, IRS passive beamforming,  device power allocation and communication scheduling. However, the formulated non-linear fractional programming problem is challenging to solve due to its non-convexity and coupled variables. To overcome the difficulty, we propose an alternating optimization based algorithm to solve it sub-optimally by leveraging the Dinkelbach's algorithm, successive convex approximation (SCA) technique, and block coordinate descent (BCD) method. Extensive simulation results show that the proposed design can significantly improve the anti-jamming performance. In particular, for the remote jammer case, the proposed design can largely shorten the flight path and thus decrease the energy consumption via the signal enhancement; while for the local jammer case, which is deemed highly challenging in  conventional systems without IRS since the retreating away strategy becomes ineffective, our proposed design even achieves a higher performance gain owing to the efficient jamming signal mitigation.

\end{abstract}


\section{Introduction}
		\label{Introduction}
Benefiting from the flexible mobility and deployment, unmanned aerial vehicles (UAVs) have been employed in numerous applications in recent years, such as remote surveillance, photography, agricultural irrigation, traffic control, and cargo transportation, etc \cite{1}. By the properly designing its trajectory, the UAV tends to obtain more dominant line-of-sight (LoS) channels and thus greatly improves the communication performance \cite{2,uav1,uav2,uav3}. In particular, exploiting the UAV as an aerial data collector for Internet-of-Things (IoT) devices is expected to bring significant benefits for transmission rate, energy consumption, and service coverage \cite{uav_iot1,uav_iot2,uav_iot3}.

However, due to the broadcast nature of wireless communications, the strong LoS channels also make UAVs more vulnerable to attacks from ground, e.g., jamming, surveillance, and eavesdropping \cite{4,TVT_cui,access_wuyang,CL_gao,anti_jammer,EE_wuyang_fix,SEE_2022,EE_duobin_rot}. In particular, the malicious jamming would severely distort the signal received at the UAV and even thus disable the UAV's data collection. To tackle this challenge, in \cite{access_wuyang}, a multi-UAV data collection system was studied where the jammers are assumed to be located randomly. Via the joint design of UAVs’ trajectories, ground nodes’ (GN) scheduling and power allocation, the minimum throughput, the average throughput, and the delay-constrained minimum throughput of all GNs are improved, respectively. In \cite{CL_gao},  turning and climbing angle-constrained are considered to improve the uplink throughput by the joint design of UAV and transmit power under malicious jamming. In particular, authors in \cite{anti_jammer} studied an anti-jamming trajectory of UAV to prevent jammers from attacking the legitimate transmission in UAV-enabled communication systems under probabilistic LoS channel. Furthermore, \cite{EE_wuyang_fix} proposed an iterative algorithm which strikes a better balance between the throughput and energy consumption under malicious jamming by the optimized UAV trajectory and thus improves the energy efficiency (EE) significantly. However, in all of the above works, the UAV's anti-jamming strategies are mostly based on retreating away from the jammer for suppressing the undesired signal, which renders not only complex trajectories but also high flight energy consumption. Even worse, considering a challenging scenario when the jammer is located nearby the legitimate terrestrial user, the above retreating away strategy becomes ineffective because when the UAV retreats away from the jammer, it also keeps away from the user, which thus degrades the reception of the desired signal as well.

Recently, intelligent reflecting surface (IRS), due to its capability of smartly configuring the wireless propagation environment, has been investigated extensively \cite{5,12,13}. Specifically, IRS comprises a large number of passive units, which can reflect the wireless signal with adjustable amplitude and/or phase shift, thus achieving the so-called passive/reflective beamforming gain. The main advantage of IRS-assisted wireless communications lies in that it considerably increases the channel capacity with low power consumption and can be flexibly deployed in the radio environment. Benefiting from the above superiority, IRS has been studied under various wireless system setups \cite{6,7,cog1,cog2,8,9,10,11}. In particular, extensive research efforts have been devoted to employing IRS for improving the transmission rate of UAV communications and/or safeguarding them from security threats. It showed that by joint design of the UAV trajectory and IRS passive beamforming, the average achievable rate of the system can be significantly improved \cite{14,15}. Facing with the threat of eavesdropping, the average secrecy rate was maximized by jointly optimizing the UAV trajectory, the transmit beamforming and the phase shift of IRS in \cite{17,20}. Furthermore, for the case suffered from jamming, it was shown that IRS can help to improve the transmission rate for mitigating the jamming signal from the malicious jammer \cite{21}.
However, it should be noted that all these works only focus on the IRS passive beamforming gain in terms of rate but ignore the benifits of IRS on UAV's energy consumption and EE, which is more practically important in UAV communication systems and thus need further study. 

Motivated by the above, in this paper, we focus on how to exploit the IRS for improving the EE of UAV in an UAV data collection system, wherein a cluster of IoT devices (IDs) tend to transmit data to the UAV in the presence of malicious jamming. The main contributions are summarized as follows.
\begin{figure}
	\centering
	\includegraphics[width=8.5cm]{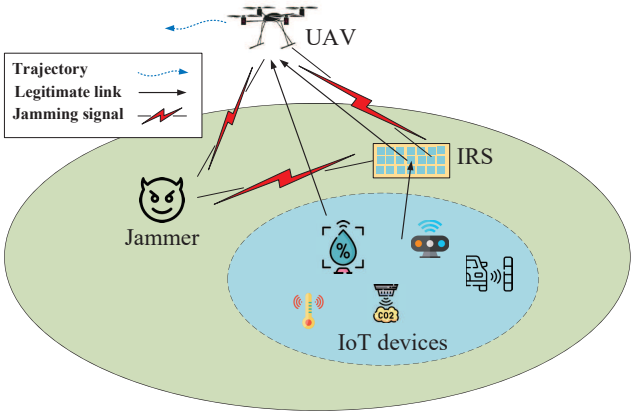}
	\caption{An IRS-assisted UAV data collection system in the presence of a jammer.} \label{Model}
			\vspace{-10pt}
\end{figure}
\vspace{-8pt}
\begin{itemize}
	\item We propose an IRS-enabled novel anti-jamming method for the UAV data collection system instead of solely relying the conventional retreating away strategy.  Specifically, by deploying an IRS nearby the IDs, the signal reception at the UAV can be significantly improved via the enhancement of the desired signal and/or the mitigation of the jamming signal, via exploiting the channel configuring capacity of the IRS. 
	
	\item We propose an alternating optimization (AO) based algorithm to solve the formulated challenging problem with coupled optimization variables and non-convex objective function. Specifically, by leveraging the block coordinate descend (BCD) method, we divide the original problem into four sub-problems that can be optimized in an iterative manner. Then, successive convex optimization (SCA) and the Dinkelbach’s algorithm are applied to settle the non-convexity of sub-problems. In particular, a low-complexity algorithm  is introduced to handle the phase shifts of the IRS. 
	
	\item Numerical results show that the proposed design can greatly improve the anti-jamming performance as compared to the benchmark ones. Specifically, two different setups corresponding to a remote  jammer (i.e., the jammer is far away from the IDs) and a local jammer (i.e., the jammer is nearby the IDs) are respectively considered for drawing some useful insights.  It is found that for the remote jammer case, the proposed design can largely shorten the flight path and thus decrease the energy consumption via the signal enhancement; while for the local jammer case when the conventional retreating away strategy becomes ineffective, our proposed design even achieves a higher performance gain owing to the efficient jamming signal mitigation.

\end{itemize}
	
\textit{Notations}: Boldface lower-case letters and boldface upper-case letters are used to denote vectors and matrices. ${\mathbb{C}^{L \times 1}}$ denotes the $L$-dimensional complex-valued vector. ${\left(  \cdot  \right)^T}$, ${\left(  \cdot  \right)^H}$ and ${\left(  \cdot  \right)^ * }$ represent the transopse, the Hermitian operations and the conjugate, respectively. $\left|  \cdot  \right|$ and $\left\|  \cdot  \right\|$ stand for the absolute value of a scalar and the Euclidean norm of a vector, respectively. ${\text{diag}}\left\{  \cdot  \right\}$ returns the diagonal matrix whose diagonals are the elements of input vector. $\Re \left(  \cdot  \right)$ and $\arg \left(  \cdot  \right)$ represent the real part and the phase of the input complex value. ${\lambda _{\max }}(\Phi)$ returns the maximum eigenvalue of $\Phi$.
	
\section{System Model and Problem Formulation}
\label{sec:System Model}
	\subsection{System Model }
	
In this paper, an IRS-aided UAV data collection system is considered as shown in Fig. \ref{Model}, where an IRS is deployed to assist in the data transmission from a cluster of IDs to a rotary-wing UAV in the presence of a jammer. The set of the IDs is denoted by $\cal K$. Without loss of generality, we consider that the UAV shares
the same frequency band for its communications and the time-division multiple access (TDMA) protocol is applied to serve the $K$ IDs. It is  assumed that both the IDs and the UAV are equipped with single omni-directional antenna, and the IRS composes of a uniform planar array (UPA) with $M=M_x \times M_z$ elements, where $M_x$ and $M_z$ denote the number of elements along the x-axis and z-axis, respectively. Then we denote the diagonal phase-shift matrix for the IRS as ${\mathbf{\Gamma }} \triangleq \left\{ {{\mathbf{\Gamma }}[n] = {\text{diag}}\left( {{e^{j{\theta _1}[n]}},{e^{j{\theta _2}[n]}},...,{e^{j{\theta _M}[n]}}} \right),\forall n} \right\}$, where ${\theta _i}[n] \in [0,2\pi ),i \in \{ 1,...,M\}$ is the phase shift of the $i$-th reflecting element in time slot $n$. All communication nodes are placed in the three dimensional (3D) Cartesian coordinates. The location of ID $k$ is represented by ${\bf{q_k}} = [{x_k},{y_k},0]$, where $k \in {\cal{K}} = \{ 1,2,...,K\}.$ The location of the jammer and IRS are denoted by ${\bf{q_j}} = [{x_j},{y_j},{z_j}]$ and ${\bf{q_r}} = [{x_r},{y_r},{z_r}]$, respectively. We assume that the location of the jammer is known by leveraging existing jammer localization techniques, such as the jamming signal strength (JSS)-based localization methods \cite{jammer-detect}. Moreover, considering the high-altitude UAV and flexibly deployed IRS, LoS channels are assumed for all the considered links.    
	
The UAV is assumed to fly at a fixed height $H_u$. The flying time of the UAV is $T$. For ease of handling, $T$ is divided into $N$ time slots, i.e., $\Delta t = T/N$, where ${\Delta t}$ is the length of one time slot. The UAV has the fixed starting point and destination, which are denoted by ${{\mathbf{q}}_{S}}$ and ${{\mathbf{q}}_{F}}$, respectively. The trajectory of the UAV can be expressed by ${\bf q_u}[n] = {[x_u[n],y_u[n],z_u[n]]^T},n \in {\cal{N}} = \{ 1,2,...,N\}$, ${\bf{Q}} \buildrel \Delta \over = \{ {\bf q_u}[n],\forall n\}$, which meets the mobility constraints as 
\begin{subequations}\label{eq0}
	\begin{spacing}{0.5}
		\begin{flalign}
			&{\bf{q}}_u\left[ 0 \right] = {{\bf{q}}_{S}},{\bf{q}}_u\left[ N \right] = {{\bf{q}}_{F}},
		\end{flalign}
	    \begin{flalign}
	    	&\left\| {{\mathbf{q}}_u[n] - {\mathbf{q}}_u[n - 1]} \right\| \leqslant {D_{\max }},n = 1,...,N,
	    \end{flalign}
    \vspace{1pt}
	\end{spacing}
\end{subequations}\!\!\!\!\!\!\!\!
where ${{D_{\max }}}$ is the maximum flying length in each time slot. Assuming that $p_k[n]$ is the transmit power of $k$-th device in time slot $n$ and ${\mathbf{P}} = \left\{ {p_k\left[ n \right],\forall n, \forall k} \right\}$, we have the following power constraints as 	
%		{\setlength\abovedisplayskip{1.5pt}
%			\setlength\belowdisplayskip{1.5pt}
\begin{equation}
	\frac{1}{N}\sum\limits_{n = 1}^N {p_k[n]}  \leqslant \bar p,\:\:
	p_k[n] \leqslant {p_{{\text{max}}}},\forall n,
\end{equation}
where $\bar p$ and ${p_{max}}$ are the average transmit power and the maximum transmit power of the IDs, respectively.
Considering the high-altitude UAV and flexibly deployed IRS, LoS channels are supposed for the ID-UAV/IRS links, the jammer-UAV/IRS links, and the IRS-UAV links. Specifically, the ID-UAV channel in time slot $n$ is expressed as
\begin{equation}
	{{\text{h}}_{ku}}[n] = \sqrt {\rho d_{ku}^{ - 2}\left[ n \right]} {e^{{{ - j2\pi {d_{ku}}\left[ n \right]} \mathord{\left/
	{\vphantom {{ - j2\pi {d_{ku}}\left[ n \right]} \lambda }} \right.
	\kern-\nulldelimiterspace} \lambda }}},
\end{equation}
where ${d_{ku}}\left[ n \right] = \left\| {{\bf{q}}_u\left[ n \right] - {{\bf{q}}_k}} \right\|$ is the distance between the ID $k$ and the UAV. $\lambda$ is the carrier wavelength and $\rho$ is the path loss at the reference distance ${D_0} = 1{\rm{m}}$. The same channel model is adopted for the channel from the jammer to the UAV, i.e., ${h_{ju}[n]}$. 
Furthermore, the reflecting channel, such as ID-IRS-UAV channel, is composed of two parts, namely, the ID-IRS channel and the IRS-UAV channel. Specifically, the IRS-UAV channel denoted by ${{\bf{h}}_{ru}}\left[ n \right] \in {\mathbb{C}}^{M \times 1}$, can be given by 
\begin{equation}
    {{\mathbf{h}}_{ru}}[n] = \sqrt {\rho d_{ru}^{ - 2}\left[ n \right]} {{\mathbf{\tilde h}}_{ru}}\left[ n \right],
\end{equation}
where ${{\mathbf{\tilde h}}_{ru}}\left[ n \right]$ denotes the phase of the channel, which is expressed by
\begin{equation}
	{{\mathbf{\tilde h}}_{ru}}\left[ n \right]={{{e}}^{{{ - j2\pi {d_{ru}}\left[ n \right]} \mathord{\left/
	{\vphantom {{ - j2\pi {d_{ru}}\left[ n \right]} \lambda }} \right.
	\kern-\nulldelimiterspace} \lambda }}}{u_x}\left[ n \right] \otimes {u_z}\left[ n \right],
\end{equation}
where
\begin{equation*}
	{u_x}\left[ n \right] \!=\! \left[ {1,{e^{ - j{{{2\pi d{\phi _{ru,x}}\left[ n \right]} \mathord{\left/{\vphantom {{2\pi d{\phi _{ru,x}}\left[ n \right]} \lambda }} \right.
	\kern-\nulldelimiterspace} \lambda }}}},...,{e^{ - j\left( {{M_x} - 1} \right){{{2\pi d{\phi _{ru,x}}\left[ n \right]} \mathord{\left/{\vphantom {{2\pi d{\phi _{ru,x}}\left[ n \right]} \lambda }} \right.
	\kern-\nulldelimiterspace} \lambda }}}}} \right]^T\!,\!
\end{equation*}
\begin{equation*}
	{u_z}\left[ n \right] \!=\! \left[ {1,{e^{ - j{{{2\pi d{\phi _{ru,z}}\left[ n \right]} \mathord{\left/{\vphantom {{2\pi d{\phi _{ru,z}}\left[ n \right]} \lambda }} \right.
	\kern-\nulldelimiterspace} \lambda }}}},...,{e^{ - j\left( {{M_z} - 1} \right){{{2\pi d{\phi _{ru,z}}\left[ n \right]} \mathord{\left/{\vphantom {{2\pi d{\phi _{ru,z}}\left[ n \right]} \lambda }} \right.
	\kern-\nulldelimiterspace} \lambda }}}}} \right]^T \!,\!
\end{equation*}
\begin{equation*}
	{{\phi _{ru,x}}}\left[ n \right] = \sin \varphi _{ru}^{\left( v \right)}\left[ n \right]\cos \varphi _{ru}^{\left( h \right)}\left[ n \right] = \frac{{{x \left[ n \right]} - {x_r}}}{{{d_{ru}\left[ n \right]}}},
\end{equation*}
\begin{equation*}
	{\phi _{ru,z}}\left[ n \right] = \sin \varphi _{ru}^{\left( v \right)}\left[ n \right]\sin \varphi _{ru}^{\left( h \right)}\left[ n \right] = \frac{{{H_u} - {z_r}}}{{{d_{ru}\left[ n \right]}}}.
\end{equation*}	
$\varphi _{ru}^{\left( v \right)}$ and $\varphi _{ru}^{\left( h \right)}$ represent the vertical and horizontal angle of departure (AoD) at the IRS, respectively.
${d_{ru}}\left[ n \right] = \left\| {{\mathbf{q}}_u\left[ n \right] - {{\mathbf{q}}_r}} \right\|$ is the distance between the IRS and the UAV. $d$ is the antenna distance.
The same model is adopted for ID-IRS channel and jammer-IRS channel. Then the channel can be expressed by 
\begin{subequations}\label{Channel_Gain}
		\begin{spacing}{0.6}
		\begin{flalign}\label{Channel_Gain_k}
			&{g_k}\left[ n \right]{\text{ = }}{\left| {{h_{ku}}\left[ n \right] + {\mathbf{h}}_{kr}^H\left[ n \right]{\mathbf{\Gamma }}\left[ n \right]{{\mathbf{h}}_{ru}}\left[ n \right]} \right|^2},		
		\end{flalign}
    	\begin{flalign}\label{Channel_Gain_j}
		&{g_j}\left[ n \right]{\text{ = }}{\left| {{h_{ju}}\left[ n \right] + {\mathbf{h}}_{jr}^H\left[ n \right]{\mathbf{\Gamma }}\left[ n \right]{{\mathbf{h}}_{ru}}\left[ n \right]} \right|^2}.		
	    \end{flalign}
	\end{spacing}
\end{subequations}

Then, the throughput during each time slot $\Delta t$ is given by 
\begin{equation}\label{Ru}
	{R_k}\left[ n \right] = \Delta tB{\log _2}\left( {1 + \frac{{{p_k}[n]{g_k}\left[ n \right]}}{{{p_j}{g_j}\left[ n \right] + {\sigma ^2}}}} \right) ,
\end{equation}
where ${p_j}$ denotes the transmit power of the jammer, ${\sigma ^2}$ denotes the power of additive white Gaussian noise (AWGN) at the UAV, and $B$ denotes the system bandwidth. Accordingly, we define a binary variable ${\mathbf{U}}{\text{ = }}\left\{ {{u_k}\left[ n \right] \in \left\{ {0,1} \right\},\forall n,k} \right\}$, which indicates whether ID $k$ is scheduled for communicating with the UAV in time slot $n$ or not, i.e., ID $k$ transmits data to the UAV if ${u_k}\left[ n \right] = 1$ and keeps silent otherwise. Moreover, we assume that only one ID is allowed to transmit data in one time slot so that the constraints can be expressed by

\begin{subequations}\label{Com_scheduled}
	\begin{spacing}{0.6}
		\begin{flalign}
			&\sum\limits_{k = 1}^K {{u_k}\left[ n \right]}  \leqslant 1,\forall n,		
		\end{flalign}
		\begin{flalign}
			&{u_k}\left[ n \right] \in \left\{ {0,1} \right\},\forall k,n.		
		\end{flalign}
	\vspace{1pt}
	\end{spacing}
\end{subequations}

In general, the UAV energy consumption is composed of two main parts. One is the communication-related energy and the other is the propulsion energy. Note that compared with 
the propulsion energy of UAVs, the communication-related energy is much smaller  (e.g., by three orders of magnitude) in practice. Thus, we only consider the propulsion energy which is required for supporting UAV's mobility and making it remain aloft during the flight period \cite{EE_zeng_rot}. The propulsion energy consumption $E_p[n]$ in Joule (J) for the rotary-wing UAV with speed $v_u\left[ n \right]$ in time slot $n$ can be modeled as 
\begin{equation}\label{Eu}
	{E_u}\left[ n \right] = \Delta t\left( {{P_0}{\vartheta _0}\left[ n \right] + {P_1}{\vartheta _1^{\frac{1}{2}}}\left[ n \right] + \frac{1}{2}d'\rho 's'Av_u^3\left[ n \right]} \right),
\end{equation}
where $d'$ and $s'$ represent fuselage drag ratio and the rotor solidity, respectively. $\rho '$ and $A$ denote the air density and the rotor disc area, respectively. $P_0$ and $P_1$ are the blade profile power and induced power which are constants related to the UAV itself. $v_u\left[ n \right]$ denotes the UAV horizontal flying speed, which is given by
\begin{equation}
	{v_u}\left[ n \right] = \frac{{\left\| {{\mathbf{q}}_u\left[ {n + 1} \right] - {\mathbf{q}}_u\left[ n \right]} \right\|}}{{\Delta t}}.
\end{equation}
$\vartheta _0\left[ n \right]$ and $\vartheta _1\left[ n \right]$ are given by 
\begin{subequations}
	\begin{spacing}{0.5}
		\begin{flalign}
		&{\vartheta _0}\left[ n \right] = 1 + \frac{{3v_u^2\left[ n \right]}}{{U_{tip}^2}},
 	    \end{flalign}
	   \begin{flalign}
	     &{\vartheta _1}\left[ n \right] = \sqrt {1 + \frac{{v_u^4\left[ n \right]}}{{v_0^4}}}  - \frac{{v_u^2\left[ n \right]}}{{2v_0^4}},
	    \end{flalign}
    \vspace{5pt}
	\end{spacing}
\end{subequations}\!\!\!\!\!\!\!
where $U_{tip}$ represents the tip speed of the UAV’s rotor blade. $v_0$ is the mean rotor induced speed when the UAV is hovering.
			
To ensure the efficient data collection, our goal is to maximize the EE of the UAV via the joint design of the UAV trajectory $\bf{Q}$, ID's transmit power $\bf{P}$, communication scheduling $\bf{U}$, and the IRS phase shift matrix ${\mathbf{\Gamma }}$. Thus, the optimization problem is formulated as 
\begin{subequations}\label{P0_ee}
	\begin{spacing}{0.6}
    	\begin{align}
			\mathop {\max }\limits_{{\mathbf{Q}},{\mathbf{P}},{\mathbf{\Gamma }}}& \;\frac{{\sum\limits_{n = 1}^N {\sum\limits_{k = 1}^K {{u_k}\left[ n \right]{R_k}\left[ n \right]} } }}{{\sum\limits_{n = 1}^N {{E_u}\left[ n \right]} }}&
		\end{align}
		\begin{align}	
			&{\rm s.t}.\: 0 \leqslant {\theta _i}\left[ n \right] \leqslant 2\pi ,\forall i,\forall n,&
		\end{align}
	\begin{align*}
		&\left( {\rm{1}} \right), \left( {\rm{2}} \right), \left( {\rm{8}} \right).&
	\end{align*}
	\end{spacing}
\end{subequations}

Problem (\ref{P0_ee}) is challenging to solve optimally since the optimization variables are coupled in the objective function with a fractional programming. However, it can be effectively solved by dividing the problem into four sub-problems by applying the BCD. Thus, this motivates us to propose an algorithm based on AO to solve problem (\ref{P0_ee}) sub-optimally, i.e., by iteratively optimizing one of sub-problems and with the other three being fixed at each iteration until convergence is achieved, as detailed in the next section.

%\section{The Proposed Alternating Algorithm}
\section{The proposed alternating algorithm}\label{A1}

\subsubsection{Transmit Power Optimization}			
%			{\setlength\abovedisplayskip{1.5pt}
%					\setlength\belowdisplayskip{1.5pt}
For given the UAV trajectory ${\bf{Q}}$, communication scheduling ${\bf{U}}$, and IRS phase shift matrix $\bf{\Gamma} $, the problem (\ref{P0_ee}) can be reduced to
\begin{subequations}\label{P_transmit}
	\begin{spacing}{0.6}
		\begin{align}
			&\mathop {\max }\limits_{\mathbf{P}} \sum\limits_{n = 1}^N\sum\limits_{k = 1}^K {{u_k}\left[ n \right]{{\log }_2}\left( {1 + \frac{{{p_k}[n]{g_k}\left[ n \right]}}{{ {{p_j}{g_j}\left[ n \right]}  + {\sigma ^2}}}} \right)} 
		\end{align}	
		\begin{align}
			\!\!\!\!\!\!\!\!{\text{s}}{\text{.t}}{\text{.}\:
			}\frac{1}{N}\sum\limits_{n = 1}^N {p[n]}  \leqslant \bar p,&
		\end{align}
		\begin{align}	
			p[n] \leqslant {p_{{\text{max}}}},\forall n.&
		\end{align}
	\end{spacing} 
\end{subequations}\!\!\!\!\!\!\!\!
This is a standard convex optimization problem that can be efficiently solved by CVX.

\subsubsection{Communication Scheduling Optimization}
For any given UAV trajectory ${\bf{Q}}$, IRS passive beamforming $\bf{\Gamma} $, and device transmit power ${\bf{P}}$, we relax the
binary variables in (\ref{P0_ee}) into continuous variables, which yields the following problem
\begin{subequations}\label{P_com}
	\begin{spacing}{0.6}
		\begin{align}
			&\mathop {\max }\limits_{\mathbf{U}} \sum\limits_{n = 1}^N\sum\limits_{k = 1}^K {{u_k}\left[ n \right]{{\log }_2}\left( {1 + \frac{{{p_k}[n]{g_k}\left[ n \right]}}{{{p_j}{g_j}\left[ n \right] + {\sigma ^2}}}} \right)} 
		\end{align}	
		\begin{align}
			\!\!\!\!\!\!\!\!{\text{s}}{\text{.t}}{\text{.}\:
			}\sum\limits_{k = 1}^K {{u_k}\left[ n \right]}  \leqslant 1,\forall n,&
		\end{align}
		\begin{align}	
			0 \leqslant {u_k}\left[ n \right] \leqslant 1,\forall n,k.&
		\end{align}
	\end{spacing} 
\end{subequations}\!\!\!\!\!\!\!\!
Note that (\ref{P_com}) is a standard linear program (LP), it can be efficiently solved by CVX. Moreover, it is easy to observed that the constraints (14b) and (14c) are met with equalities when the optimal solution $\bf U$ is attained.

\subsubsection{IRS Passive Beamforming Optimization}
To solve the subproblem with given any trajectory {\bf Q}, communication scheduling {\bf U}, and transmit power {\bf P}, the optimal phase shifts of IRS can be obtained by solving
\begin{subequations}\label{P2}
	\begin{spacing}{0.6}
		\begin{align}
			&\mathop {\max }\limits_{\mathbf{\Gamma }} \frac{{{\text{ }}{{\left| {{h_{ku}}\left[ n \right] + {\mathbf{h}}_{kr}^H\left[ n \right]{\mathbf{\Gamma }}\left[ n \right]{{\mathbf{h}}_{ru}}\left[ n \right]} \right|}^2}}}{{{{\left| {{h_{ju}}\left[ n \right] + {\mathbf{h}}_{jr}^H\left[ n \right]{\mathbf{\Gamma }}\left[ n \right]{{\mathbf{h}}_{ru}}\left[ n \right]} \right|}^2}}}
		\end{align}
		\begin{align}
			{\text{s}}{\text{.t}}{\text{.}}\;{\theta _i}\left[ n \right] \in \left[ {0,2\pi } \right),\forall i,\forall n.
		\end{align}	
	\end{spacing} 
\end{subequations}\!\!\!\!\!\!\!\!
It is quite hard to obtain the optimal solution due to the unit modulus constraints. By invoking the equality ${{\mathbf{a}}^H}{\mathbf{\Gamma b}} = {{\mathbf{v}}^H}{\text{diag}}\left\{ {{{\mathbf{a}}^H}} \right\}{\mathbf{b}}$, where $v[n] = [{e^{j{\theta _1}[n]}},{e^{j{\theta _2}[n]}},...,{e^{j{\theta _M}[n]}}]$, (\ref{Channel_Gain_k}), (\ref{Channel_Gain_j}) can be rewritten as 
\begin{equation}
	{g_k}\left[ n \right]{\text{ = }}{\left| {{\mathbf{v}}{{[n]}^H}{{\bf a} _k}[n] + {{\tilde {\bf a} }_k}[n]} \right|^2},
\end{equation}
\begin{equation}
	{g_j}\left[ n \right]{\text{ = }}{\left| {{\mathbf{v}}{{[n]}^H}{{\bf a} _j}[n] + {{\tilde {\bf a} }_j}[n]} \right|^2},
\end{equation}
where ${{{\bf a}}_k}[n] = {\text{diag}}\left\{ {{\mathbf{h}}_{kr}^H[n]} \right\}{{\mathbf{h}}_{ru}}[n]\sqrt {p_k[n]} $, ${{{\bf a}}_j}[n] = {\text{diag}}\left\{ {{\mathbf{h}}_{jr}^H[n]} \right\}{{\mathbf{h}}_{ru}}[n]\sqrt {{p_j}} $,
${\tilde {\bf a} _k}[n] = h_{ku}^H[n]\sqrt {p_k[n]}$, ${\tilde {\bf a} _j}[n] = h_{ju}^H[n]\sqrt {{p_j}} $.
The problem is non-convex and is not easy to solve directly. As such, problem (\ref{P2}) in each time slot can be equivalently written as
\begin{subequations}\label{P2.1}
	\begin{spacing}{0.6}
		\begin{align}
		\mathop {\max }\limits_{\mathbf{v}} &\frac{{ {{{\left| {{\mathbf{v}}{{[n]}^H}{{\bf a} _k}[n] + {{\tilde {\bf a} }_k}[n]} \right|}^2}} }}{{ {{{\left| {{\mathbf{v}}{{[n]}^H}{{\bf a} _j}[n] + {{\tilde {\bf a} }_j}[n]} \right|}^2}}  + {\sigma ^2}}},\forall n 
	 \end{align}
 		\begin{align}
 		{\text{s}}{\text{.t}}{\text{.}}\left| {{v_i}[n]} \right| = 1, \forall i, \forall n. 
 	\end{align}
\end{spacing} 
\end{subequations}\!\!\!\!\!\!\!\!
By assuming the transmit power of ID $k$ in time slot $n$ is not zero, problem (\ref{P2.1})  can be equivalently transformed into 
\begin{subequations}\label{P2.2}
	\begin{spacing}{0.6}
		\begin{align}
			\mathop {\min }\limits_{\mathbf{v}} &\frac{{ {{{\left| {{\mathbf{v}}{{[n]}^H}{{\bf a} _j}[n] + {{\tilde {\bf a} }_j}[n]} \right|}^2}}  + {\sigma ^2}}}{{ {{{\left| {{\mathbf{v}}{{[n]}^H}{{\bf a} _k}[n] + {{\tilde {\bf a} }_k}[n]} \right|}^2}} }},\forall n
		\end{align}
		\begin{align}
			{\text{s}}{\text{.t}}{\text{.}}\left| {{v_i}[n]} \right| = 1, \forall i, \forall n. 
		\end{align}
	\end{spacing} 
\end{subequations}\!\!\!\!\!\!\!\!
The problem above belongs to fractional programming and can be converted into a series of parametric sub-problems. In the $i$-th iteration, we consider the following parametric problem. 
\begin{subequations}\label{P2.3}
	\begin{spacing}{0.6}
		\begin{align}
	         \mathop {\min }\limits_{\mathbf{v}} {\left| {{\mathbf{v}}{{[n]}^H}{{\bf a} _j}[n] \!+\! {{\tilde {\bf a} }_j}[n]} \right|^2} \!+\! {\sigma ^2}  \!-\! {\eta ^{(i - 1)}}{\left| {{\mathbf{v}}{{[n]}^H}{{\bf a} _k}[n] \!+\! {{\tilde {\bf a} }_k}[n]} \right|^2}
		\end{align}
	\begin{align}
		{\text{s}}{\text{.t}}{\text{.}}\left| {{v_i}[n]} \right| = 1, \forall i, \forall n. 
    \end{align}
\end{spacing} 
\end{subequations}\!\!\!\!\!\!\!\!
where $\eta^{(i-1)}$ is an introduced parameter with an initial value $\eta^{(0)}=0$. $\eta^{(i)}$ can be updated by
\begin{equation}
	{\eta ^{\left( i \right)}} = \frac{{ {{{\left| {{{\mathbf{v}}^{\left( i \right)H}}[n]{{\bf a} _j}[n] + {{\tilde {\bf a} }_j}[n]} \right|}^2}}  + {\sigma ^2}}}{{ {{{\left| {{{\mathbf{v}}^{\left( i \right)H}}{{[n]}^H}{{\bf a} _k}[n] + {{\tilde {\bf a} }_k}[n]} \right|}^2}} }},
\end{equation}
where ${{{\mathbf{v}}^{\left( i \right)H}}[n]}$ can be obtained by solving problem (\ref{P2.3}). To obtain the optimal value of $\bf{v}$, we transform the objective function and minimize an upper bound as follows \cite{23}
\begin{subequations}
		\begin{align*}
        	&{\left| {{{\mathbf{v}}^H}[n]{{\bf a} _j}[n] + {{\tilde {\bf a} }_j}[n]} \right|^2} + {\sigma ^2} - \eta {\left| {{{\mathbf{v}}^H}{{[n]}^H}{{\bf a} _k}[n] + {{\tilde {\bf a} }_k}[n]} \right|^2}  &
        \end{align*}
	    \begin{align*}
	      & = {{\mathbf{v}}^H}\left( {{{\mathbf{{\bf a} }}_j}{\mathbf{{\bf a} }}_j^H - \eta {{\mathbf{{\bf a} }}_k}{\mathbf{{\bf a} }}_k^H} \right){\mathbf{v}} + {\left| {{{\tilde {\bf a} }_j}} \right|^2} + {\sigma ^2}   &
        \end{align*}
	    \begin{align*}
	     \;\;\;\; & - \eta {\left| {{{\tilde {\bf a} }_k}} \right|^2} - 2\Re \left\{ {{{\mathbf{v}}^H}\left( {\eta \tilde {\bf a} _k^*{{{\bf a}}_k} - \tilde \alpha _j^*{{{\bf a}}_j}} \right)} \right\} &
        \end{align*}
	    \begin{align}
	      &\leqslant{\lambda _{\max }}({\mathbf{\Phi }}){\left\| {\mathbf{v}} \right\|^2} - 2\Re \left\{ {{{\mathbf{v}}^H}{\mathbf{\omega }}} \right\} + c, &
        \end{align} 
\end{subequations}
where 
\begin{subequations}
	    \begin{align}
	       &\!\!\!\!\!\!\!\!\!\!\!\!\!\!\!\!\!\!\!\!\!\!\!\!\!\!\!\!\!\!\!\!\!\!\!\!\!\!\!\!\!\!{\bf a}_j=[{\bf a}_j[1],...,{\bf a}_j[N]],&
	    \end{align}
        \begin{align}	       	      
	        &\!\!\!\!\!\!\!\!\!\!\!\!\!\!\!\!\!\!\!\!\!\!\!\!\!\!\!\!\!\!\!\!\!\!\!\!\!\!{\bf a}_k=[{\bf a}_k[1],...,{\bf a}_k[N]], \forall k, &
	    \end{align}
		\begin{align}
	        &\!\!\!\!\!\!\!\!\!\!\!\!\!\!\!\!\!\!\!\!\!\!\!\!\!\!\!\!\!\!\!\!\!\!\!\!\!\!\!\!\!\!\!\!{\mathbf{\Phi }} = {{\bf a}_j}{{\bf a}}_j^H - \eta {{\bf a}_k}{{\bf a}}_k^H, &
	   \end{align}
	   \begin{align}\label{omiga}
	       &\!\!\!\!\!\!\!\!\!\!\!\!{\mathbf{\omega }} = \left( {{\lambda _{\max }}\left( {\mathbf{\Phi }} \right){\mathbf{I}} - \Phi } \right){\mathbf{\tilde v}} + \eta \tilde {\bf a} _k^*{{{\bf a}}_k} - \tilde {\bf a} _j^*{{{\bf a}}_j} , &
	   \end{align}
	   \begin{align}
	       &c = {{\mathbf{\tilde v}}^H}\left( {{\lambda _{\max }}\left( {\mathbf{\Phi }} \right){\mathbf{I}} - \Phi } \right){\mathbf{\tilde v}} + {\left| {{{\tilde {\bf a} }_j}} \right|^2} + {\sigma ^2} - \eta {\left| {{{\tilde {\bf a} }_k}} \right|^2} ,&
       \end{align}
\end{subequations}
where $\tilde v$ is the previous iteration of the solution to $v$. Therefore, the final optimization problem is given by 
\begin{subequations}
	\begin{spacing}{0.6}
		\begin{align}
	         \min \;\;{\lambda _{\max }}({\mathbf{\Phi }}){\left\| {\mathbf{v}} \right\|^2} - 2\Re \left\{ {{{\mathbf{v}}^H}{\mathbf{\omega }}} \right\} 
        \end{align}
	    \begin{align}
	         {\text{s}}{\text{.t}}{\text{.}}\left| {{v_i}\left[ n \right]} \right| = 1, \forall i,\forall n.
        \end{align}
\end{spacing} 
\end{subequations}\!\!\!\!\!\!\!\!
It is obvious that the optimal value is obtained when $\Re \left\{ {{{\mathbf{v}}^H}{\mathbf{\omega }}} \right\}$ is maximized, i.e., the phase of ${v_i}$ and ${\omega _i}$ are equal, where ${\omega _i}$ is the $i$-th element of ${\mathbf{\omega }}$. Accordingly, we obtain the final solution in the time slot $n$ as 
\begin{eqnarray}
	{\mathbf{ v}}\left[ n \right] = {\left[ {{e^{-j\arg \left( {{\omega _1}} \right)}},...,{e^{-j\arg \left( {{\omega _M}} \right)}}} \right]^T}.
\end{eqnarray} 
Finally, we obtain a closed-form solution in each iteration and the computational complexity is significantly reduced.

\begin{algorithm}[t]
	\renewcommand{\algorithmicrequire}{\textbf{Input:}}
	\renewcommand{\algorithmicensure}{\textbf{Output:}}
	%\caption{Joint trajectory and scheduling optimization algorithm}\label{Algorithm1}
	\caption{An alternating algorithm for solving {(\ref{P2}})}\label{Algorithm1}
	\begin{algorithmic} [1]
		\STATE\textbf{Input:}  devicess transmit power ${\bf P}$, communication sheduling ${\bf U}$, and UAV trajectory ${\bf Q}$.
		\STATE{Initialization:}~set~initial $\mu_1$, $i_{max}$, and $\eta^{(0)}$.
		~Set~$i = 0$, as iteration index, ${\bf{\tilde \Gamma}}$ as initial IRS phase shift. 
		\STATE\textbf{~repeat.}
		\STATE~ Fix $\bf \Gamma$ and calculate $\eta^{(i)}$ using (19) and ${\bf \omega}$ using (21).	
		\STATE~ Set ${\mathbf{\tilde \Gamma }}{\text{ = diag}}\left\{ {{e^{ - j{\mathbf \omega}}}} \right\}$.
		\STATE~ {Update $ i \leftarrow i + 1\ $}.
		\STATE\textbf{~until}: convergence or $i = {i_{\max }}$. 	
		\STATE\textbf{Output:}${\bf \Gamma}$. 
	\end{algorithmic}
\end{algorithm}
\subsubsection{UAV Trajectory Optimization}

For given transmit power ${\bf{P}}$, communication scheduling ${\bf{U}}$, and phase shift $\bf{\Gamma} $, (\ref{P0_ee}) can be rewritten by
\begin{subequations}\label{P3}
	\begin{spacing}{0.6}
		\begin{align}
			\mathop {\max }\limits_{\mathbf{Q}} \frac{{B\sum\limits_{n \in \cal N} {\sum\limits_{k \in \cal K} {\Delta t{{\log }_2}\left( {1 + \frac{{{p_k}[n]{g_k}\left[ n \right]}}{{ {{p_j}{g_j}\left[ n \right]}  + {\sigma ^2}}}} \right)} } }}{{\sum\limits_{n \in \cal N} {\Delta t\left( {{P_0}{\vartheta _0}\left[ n \right] + {P_1}\vartheta _1^{\frac{1}{2}}\left[ n \right] + \frac{1}{2}d'\rho 's'Av_u^3\left[ n \right]} \right)} }}
		\end{align}
		\begin{align}
			{\text{s}}{\text{.t}}{\text{.}}\:&{\bf{q}}_u\left[ 0 \right] = {{\bf{q}}_{S}},{\bf{q}}_u\left[ N \right] = {{\bf{q}}_{F}},&
		\end{align}
		\begin{align}
			\:\:\:\:\:\:\:\:\:\:\:\:&\left\| {{\mathbf{q}}_u[n] - {\mathbf{q}}_u[n - 1]} \right\| \leqslant {D_{\max }},n = 1,...,N.&
		\end{align}	
	\end{spacing}
\end{subequations}\!\!\!\!\!\!\!\!\!
Problem (\ref{P3}) is difficult to solve due to its non-convex objective function with a fractional objective function. Note that ${{{\mathbf{\tilde h}}}_{ru}}\left[ n \right]$, ${{{\mathbf{\tilde h}}}_{ku}}\left[ n \right]$ and ${{{\mathbf{\tilde h}}}_{ju}}\left[ n \right]$ are relative to the trajectory variables. However, it is observed that those are more complex to handle due to their non-linear formation. To overcome such a problem, we use ($i-1$)th iteration to obtain the approximate ${{{\mathbf{\tilde h}}}_{ru}}\left[ n \right]$,${{{\mathbf{\tilde h}}}_{ku}}\left[ n \right]$ and ${{{\mathbf{\tilde h}}}_{ju}}\left[ n \right]$ \cite{20}. Based on the consideration above, we consider rewriting the objective function. By denoting
\[\begin{gathered}
	{{\bf{H}}_k}\left[ n \right] = \left[ {\sqrt \rho  \tilde h_{ku}^{\left(i-1\right)}\left[ n \right],\sqrt \rho  {\mathbf{h}}_{kr}^H\left[ n \right]{\mathbf{\Gamma }}\left[ n \right]{{{\mathbf{\tilde h}}}_{ru}^{\left(i-1\right)}}\left[ n \right]} \right], \hfill \\
	{{\bf{H}}_j}\left[ n \right] = \left[ {\sqrt \rho  \tilde h_{ju}^{\left(i-1\right)}\left[ n \right],\sqrt \rho  {\mathbf{ h}}_{jr}^H\left[ n \right]{\mathbf{\Gamma }}\left[ n \right]{{{\mathbf{\tilde h}}}_{ru}^{\left(i-1\right)}}\left[ n \right]} \right], \hfill \\ 
\end{gathered} \]
where  
${\tilde h_{ku}}[n] = {e^{ - j\frac{{2\pi {d_{ku}}\left[ n \right]}}{\lambda }}}$, (\ref{Channel_Gain}) can be transformed as				
\begin{subequations}
		\begin{align}
			{g_k}\left[ n \right] = {\mathbf{d}}_k^T[n]{\mathbf{H}}_k^H[n]{{\mathbf{H}}_k}[n]{{\mathbf{d}}_k}[n],	
		\end{align}
		\begin{align}	
			{g_j}\left[ n \right] = {\mathbf{d}}_j^T[n]{\mathbf{H}}_j^H[n]{{\mathbf{H}}_j}[n]{{\mathbf{d}}_j}[n],
		\end{align}
\end{subequations}
where ${{\mathbf{d}}_k}[n] \!=\! {\left[ {{ {{d_{ku}^{-1}}[n]} } ,{{{d_{ru}^{ - 1}}[n]} } } \right]^T}$, ${{\mathbf{d}}_j}[n] = {\text{ }}{\left[ {{{{d_{ju}^{ - 1}}[n]}} , {{{d_{ru}^{ - 1}}[n]}} } \right]^T}$.

Moreover, the other problem is the non-convex objective function which is composed of the coupled variables. To transform the numerator of the objective function of (\ref{P3}) into concave, we consider introducing the slack variables and leveraging the SCA method. By introducing the slack variables  ${\mathbf{S}} = \{ S_k[n],\forall n,k\}$ and ${\mathbf{G}} = \{ G[n],\forall n\}$, the throughput of ID $k$ in time slot $n$, i.e., ${R_k}\left[ n \right]$, can be rewritten as follows
\begin{equation}\label{Ru_lower}
	{R_k}\left[ n \right] = \Delta tB{\log _2}\left( {1 + \frac{{{S^{ - 1}_k}\left[ n \right]}}{{G\left[ n \right]}}} \right),\forall n,k,
\end{equation}
with two additional constraints
\begin{equation}\label{C1}
	p_k[n]{\mathbf{d}}_k^T[n]{\mathbf{H}}_k^H[n]{{\mathbf{H}}_k}[n]{{\mathbf{d}}_k}[n] \geqslant {S^{ - 1}}[n],\forall n, k,
\end{equation} 
and
\begin{equation}\label{C2}  
	{p_j}{\mathbf{d}}_j^T[n]{\mathbf{H}}_j^H[n]{{\mathbf{H}}_j}[n]{{\mathbf{d}}_j}[n] + {\sigma ^2} \leqslant G[n],\forall n. 
\end{equation}
The equivalence has been proved by contradiction in \cite{access_wuyang}. Specifically, if the constraints hold with inequalities, we can improve the value of the objective function by decreasing variable ${\mathbf{S}}$ and ${\mathbf{G}}$.

In the next step, we focus on dealing with problem (\ref{Ru_lower}). Note that constraints (\ref{C1}) and (\ref{C2}) are still non-convex, thus, with the property of the convex function's first-order Taylor expansion and successive convex optimization method, we first use the following Lemma to obtain a lower bound of the objective function (\ref{P3}).

\begin{lemma}\label{lemma2}
	The numerator of the objective function (\ref{P3}) is lower bounded at the feasible point $ (S_{k,0}[n],{G_0}[n])\ $ by
	\begin{spacing}{1.0}
		\begin{equation}\label{eq24}
			\begin{array}{l}
				\tilde R_k\left[ n \right] = {\log_2}\left( {1 + \frac{1}{{{S_{k,0}}[n]{G_0}[n]}}} \right)  
				+ {\zeta _{k,1}}\left[ n \right]\left( {S_k\left[ n \right] - {S_{k,0}}\left[ n \right]} \right)\\ + {\zeta _{k,2}}\left[ n \right]\left( {G\left[ n \right] - {G_0}\left[ n \right]} \right), \forall n,k,
			\end{array}\
		\end{equation}
	\end{spacing}
	\noindent where ${\zeta _{k,1}}[n] =  - {\log _2}\left( {\frac{e}{{{S_{k,0}}[n] + {{({S_{k,0}}[n])}^2}{G_0}[n]}}} \right)$, ${\zeta _{k,2}}[n] =  - {\log _2}\left( {\frac{e}{{{G_0}[n] + {{({G_0}[n])}^2}{S_{k,0}}[n]}}} \right)$. 
\end{lemma}
\vspace{7pt}

\indent \emph{Proof:}
	Since $f(x,y) = {\log _2}(1 + 1/xy)$ is a convex function \cite{hai_TWC}, its first-order Taylor expansion provides a global under-estimator at a feasible point $({x_0},{y_0})$, i.e.,
	\begin{equation}\label{eq25}
		\begin{array}{l}
			{\log _2}(1 + 1/xy) \ge {\log _2}(1 + 1/{x_0}{y_0})\\
			- (x - {x_0}){\log _2}e/({x_0} + {({x_0})^2}{y_0})\\
			- (y - {y_0}){\log _2}e/({y_0} + {({y_0})^2}{x_0}).
		\end{array}\
	\end{equation}
	Thus, by applying $ x = {S_{k}}[n]\ $, $ y = G[n]\ $, \textbf{ \emph{Lemma}} \emph{1} is proved.

Therefore, we can obtain a lower bound of objective function.
However, the constraints (\ref{C1}) and (\ref{C2}) are non-convex because of the coupled variables. To deal with the non-convexity, we introduce the slack variables ${{\bf{\xi }}_1} = \left\{ {{\xi _1}\left[ n \right],\forall n} \right\}$, ${{\mathbf{\xi }}_2} = \left\{ {{\xi _2}\left[ n \right],\forall n} \right\}$, ${{\mathbf{\xi }}_3} = \left\{ {{\xi _3}\left[ n \right],\forall n} \right\}$, ${{\mathbf{\xi }}_4} = \left\{ {{\xi _4}\left[ n \right],\forall n} \right\}$, and by denoting ${{\mathbf{\tilde d}}_k} = {\left[ {{\xi _1}\left[ n \right],{\xi _2}\left[ n \right]} \right]^T}$, ${{\mathbf{\tilde d}}_j} = {\left[ {{\xi _3}\left[ n \right],{\xi _4}\left[ n \right]} \right]^T}$, the constraints (\ref{C1}) and (\ref{C2}) are rewritten as
\begin{equation}
	p_k[n]\tilde {\bf{d}}_{k}^T[n]{\bf{H}}_{k}^H[n]{{\bf{H}}_{k}}[n]{{\tilde {\bf{d}}}_{k}}[n] \ge {S^{ - 1}_k}[n], \forall n,k\\
\end{equation}
and
\begin{equation}\label{jammer_power}	
	{p_j}\tilde {\bf{d}}_{j}^T[n]{\bf{H}}_{j}^H[n]{{\bf{H}}_{j}}[n]{{\tilde {\bf{d}}}_{j}}[n] + {\sigma ^2}\le G[n], \forall n\\
\end{equation}
with for additional constraints
\begin{subequations}
	\begin{spacing}{0.45}
		\begin{align}
			&{d_{ku}^{ -1 }}[n]  \ge {\xi _1}[n],\forall n, \forall k,
		\end{align}
		\begin{align}
			&{d_{ru}^{ -1 }}[n]  \ge  {\xi _2}[n],\forall n,
		\end{align}
		\begin{align}
			& {d_{ju}^{ -1 }[n]}  \le  {\xi _3}[n],\forall n, 
		\end{align}	
		\begin{align}
			&{d_{ru}^{ -1 }}[n]  \le {\xi _4}[n],\forall n.
		\end{align}
		\begin{align*} 
		\end{align*}
	\end{spacing}
\end{subequations}
In order to conveniently tackle the non-convex variables, we unfold as follows
\begin{subequations}\label{F_krj}
	\begin{spacing}{0.45}
		\begin{align}
			{F_k}\left[ n \right] - {{{\xi _1}^{ - 2}\left[ n \right]}} \leqslant 0,\forall n,k,  
		\end{align}	
		\begin{align}
			{F_r}\left[ n \right] - { {{\xi _2}^{ - 2}\left[ n \right]} } \leqslant 0,\forall n, 
		\end{align}	
		\begin{align}
			{{\xi _3}^{ - 2}\left[ n \right]}  - {F_j}\left[ n \right] \leqslant 0,\forall n,  
		\end{align}	
		\begin{align}
			{{\xi _4}^{ - 2}\left[ n \right]}  - {F_r}\left[ n \right] \leqslant 0,\forall n, 
		\end{align}       
		\begin{align*} 
		\end{align*}
	\end{spacing}
\end{subequations}\!\!\!\!\!\!\!\!\!
where ${F_k}\left[ n \right] = {\left( {x\left[ n \right] - {x_k}} \right)^2} + {\left( {y\left[ n \right] - {y_k}} \right)^2} + {H_u^2}$, ${F_r}\left[ n \right] = {\left( {x\left[ n \right] - {x_r}} \right)^2} + {\left( {y\left[ n \right] - {y_r}} \right)^2} + {\left( {H_u - {z_r}} \right)^2}$, ${F_j}\left[ n \right] = {\left( {x\left[ n \right] - {x_j}} \right)^2} + {\left( {y\left[ n \right] - {y_j}} \right)^2} + {\left( {H_u - {z_j}} \right)^2}$. However, there still exists several non-convex feasible regions in constraint (\ref{F_krj}). We use SCA technique to address the non-convex regions. The first-order Taylor expansions of ${ {{\xi _1}^{ - 2}\left[ n \right]}}$, ${{{\xi _2}^{ - 2}\left[ n \right]} }$, ${\mathbf{d}}_k^T[n]{\mathbf{H}}_k^H[n]{{\mathbf{H}}_k}[n]{{\mathbf{d}}_k}[n]$ at the feasible points ${{\mathbf{\xi }}_{{\mathbf{1}},{\mathbf{0}}}} = \left\{ {{\xi _{1,0}}\left[ n \right],\forall n} \right\}$, ${{\mathbf{\xi }}_{{\mathbf{2}},{\mathbf{0}}}} = \left\{ {{\xi _{2,0}}\left[ n \right],\forall n} \right\}$ and ${{\mathbf{\tilde d}}_{k,0}} = \left\{ {{{\tilde d}_{k,0}}[n],\forall n} \right\}$, which is given by
\begin{subequations}
	\begin{spacing}{0.6}
		\begin{align}
			&\xi _1^{ - 2}\left[ n \right] \geqslant \xi _{1,0}^{ - 2}\left[ n \right] - 2\xi _{1,0}^{ - 2 - 1}\left[ n \right]\left( {{\xi _1}\left[ n \right] - {\xi _{1,0}}\left[ n \right]} \right),\\\
			&\xi _2^{ - 2}\left[ n \right] \geqslant \xi _{2,0}^{ - 2}\left[ n \right] - 2\xi _{2,0}^{ - 2 - 1}\left[ n \right]\left( {{\xi _2}\left[ n \right] - {\xi _{2,0}}\left[ n \right]} \right),
		\end{align}	
		\begin{align*}	
			&{\mathbf{\tilde d}}_k^T[n]{\mathbf{H}}_k^H[n]{{\mathbf{H}}_k}[n]{{{\mathbf{\tilde d}}}_k}[n] \geqslant  - {\mathbf{\tilde d}}_{k,0}^T[n]{\mathbf{H}}_k^H[n]{{\mathbf{H}}_k}[n]{{{\mathbf{\tilde d}}}_{k,0}}[n] 
		\end{align*}	
		\begin{align}
			&+ 2\Re \left[ {{\mathbf{\tilde d}}_{k,0}^T[n]{\mathbf{H}}_k^H[n]{{\mathbf{H}}_{\text{k}}}[n]{{{\mathbf{\tilde d}}}_k}[n]} \right]. 
		\end{align} 
	\end{spacing}
\end{subequations} 
To make the problem solvable, we use the first-order Taylor expansion of the convex function to obtain the lower bound. At the given feasible points ${{\mathbf{x}}_0} = \left\{ {{x_0}\left[ n \right],\forall n} \right\}$ and ${{\mathbf{y}}_0} = \left\{ {{y_0}\left[ n \right],\forall n} \right\}$, the first-order Taylor expansions are given by 					
\begin{subequations}\label{xy_talor}
	\begin{spacing}{0.45}
		\begin{align}
			{x^2}\left[ n \right]\geqslant - x_0^2\left[ n \right] + {x_0}\left[ n \right]x\left[ n \right], 
		\end{align}	 
		\begin{align}	
			{y^2}\left[ n \right] \geqslant- y_0^2\left[ n \right] + {y_0}\left[ n \right]y\left[ n \right].
		\end{align}	
    \vspace{5pt}
	\end{spacing} 
\end{subequations} 

As such, ${F_r}\left[ n \right]$ and ${F_j}\left[ n \right]$  can be transformed as
\begin{equation}
	\begin{gathered}
		{{\tilde F}_r}\left[ n \right] =  - x_0^2\left[ n \right] + 2{x_0}\left[ n \right]x\left[ n \right] + x_r^2 - y_0^2\left[ n \right] + \\ 2{y_0}\left[ n \right]y\left[ n \right]-y_r^2 \!+\! H_u^2-2{x_r}x\left[ n \right] - 2{y_r}y\left[ n \right] - 2{z_r}H_u. \hfill \\ 
	\end{gathered}
\end{equation}
\begin{equation}
	\begin{gathered}
		{{\tilde F}_j}\left[ n \right] =  - x_0^2\left[ n \right] + 2{x_0}\left[ n \right]x\left[ n \right] + x_j^2 - y_0^2\left[ n \right] + \\ 2{y_0}\left[ n \right]y\left[ n \right]-y_j^2 \!+\! H_u^2-2{x_j}x\left[ n \right] - 2{y_j}y\left[ n \right] - 2{z_j}H_u. \hfill \\ 
	\end{gathered}
\end{equation}
Thus (\ref{C1}) can be writen as 
\begin{equation}\label{GN_power}
	\begin{aligned}
		p_k[n]&(2\Re \left[ {{\mathbf{\tilde d}}_{g,0}^T[n]{\mathbf{H}}_g^H[n]{{\mathbf{H}}_{\text{g}}}[n]{{{\mathbf{\tilde d}}}_g}[n]} \right]-\\
		&{{\tilde{\bf{d}}}_{g,0}}^T[n]{\bf{H}}_{g}^H[n]{{\bf{H}}_{g}}[n]{{\tilde {\bf{d}}}_{g,0}}[n]) \ge {S^{ - 1}}[n], \forall n,k.
	\end{aligned}
\end{equation}
Further more, by denoting
\begin{subequations}
	\begin{spacing}{0.6}
		\begin{align}
			&{ {{{\tilde \xi }_1}\left[ n \right]} } =  {\xi _{1,0}^{ - 2}\left[ n \right]}  - 2{ {\xi _{1,0}^{ -3}\left[ n \right]} }\left( {{\xi _1}\left[ n \right] - {\xi _{1,0}}\left[ n \right]} \right),
		\end{align}			
		\begin{align} 
			&{ {{{\tilde \xi }_2}\left[ n \right]}} = { {\xi _{2,0}^{ - 2}\left[ n \right]} } - 2{ {{\xi _{2,0}^{ - 3}}\left[ n \right]} }\left( {{\xi _2}\left[ n \right] - {\xi _{2,0}}\left[ n \right]} \right),
		\end{align}
	\vspace{1pt}
	\end{spacing}	
\end{subequations}\!\!\!\!\!\!\!\!
we can substitute all concave points to solvable ones.
Then, to deal with the non-convexity of $E_u[n]$ in the objective function of problem (\ref{P3}), we introduce the other slack variable ${\mathbf{e}} = \left\{ {e\left[ n \right],\forall n} \right\}$ such that
\begin{equation}\label{eq_k}
	e\left[ n \right] \geqslant {\left( {\sqrt {1 + \frac{{v_u^4\left[ n \right]}}{{v_0^4}}}  - \frac{{v_u^2\left[ n \right]}}{{2v_0^4}}} \right)^{\frac{1}{2}}}.
\end{equation}
Next, we can rewrite (\ref{eq_k}) as 
\begin{equation}\label{eq_re_k}
	\frac{1}{{{e^2}\left[ n \right]}} \leqslant {e^2}\left[ n \right] + \frac{{v_u^2}}{{v_0^2}} = {e^2}\left[ n \right] + \frac{{{{\left\| {{\mathbf{q}}_u\left[ {n + 1} \right] - {\mathbf{q}}_u\left[ n \right]} \right\|}^2}}}{{v_0^2\Delta {t^2}}}.
\end{equation}
Note that the constraint (\ref{eq_re_k}) should hold with equality when the optimal solution is obtained. If the constraint (\ref{eq_re_k}) holds with inequality, we can improve the value of the objective function (\ref{P3}) by decreasing variable $e[n]$. Since that both sides of constraint (\ref{eq_re_k}) are convex, we should transform the right-hand side into concave formations. By applying the first-order Taylor expansion at given points ${{\mathbf{s}}_0} = \left\{ {{s_0}\left[ n \right],\forall n} \right\}$ and ${{\mathbf{Q}}_0} = \left\{ {{{\mathbf q}_0}\left[ n \right],\forall n} \right\}$, 
\begin{equation}
	\begin{gathered}
		{e^2}\left[ n \right] + \frac{{{{\left\| {{\mathbf{q}}_u\left[ {n + 1} \right] - {\mathbf{q}}_u\left[ n \right]} \right\|}^2}}}{{v_0^2\Delta {t^2}}} \hfill \\
		\geqslant e_0^2\left[ n \right] + 2{e_0}\left[ n \right]\left( {e\left[ n \right] - {e_0}\left[ n \right]} \right) - \frac{{{{\left\| {{{\mathbf{q}}_0}\left[ {n + 1} \right] - {{\mathbf{q}}_0}\left[ n \right]} \right\|}^2}}}{{v_0^2\Delta {t^2}}} \hfill \\
		+ \frac{2}{{v_0^2\Delta {t^2}}}{\left( {{{\mathbf{q}}_0}\left[ {n + 1} \right] \!-\! {{\mathbf{q}}_0}\left[ n \right]} \right)^T}\left( {{\mathbf{q}}_u\left[ {n + 1} \right] \!-\! {\mathbf{q}}_u\left[ n \right]} \right) \triangleq {{\cal L}_0}\left[ n \right]. \hfill \\ 
	\end{gathered} 
\end{equation}
As such, all the constraints are convex. Problem (\ref{P3}) is transformed into
\begin{subequations}\label{Final}
	\begin{spacing}{0.7}
		\begin{align}
			\mathop {\max }\limits_{
				\begin{subarray}
				     	{\mathbf{Q}}{\mathbf{Q}},{\mathbf{S}},{\mathbf{G}},{\mathbf{e}},\\
					{\mathbf{\xi_1}},{\mathbf{\xi_2}},{\mathbf{\xi_3}},{\mathbf{\xi_4}}
				\end{subarray}
				 } \:\frac{{\Delta tB\sum {_{n = 1}^N} \tilde R_u[n]}}{{\sum {_{n = 1}^N} \Delta t\left( {{P_0}{\vartheta _0}\left[ n \right] + {P_1}e\left[ n \right] + \frac{1}{2}d'\rho 's'Av_u^3\left[ n \right]} \right)}}
		\end{align}
		\begin{align}
			{\text{s}}{\text{.t}}{\text{.}}\:&\frac{1}{{{e^2}\left[ n \right]}} \leqslant {{\cal L}_0}\left[ n \right], \forall n,&
		\end{align}
		\begin{align}
			&e\left[ n \right] > 0, \forall n,&
		\end{align}	
		\begin{align}
			\:\:\:\:\:\:\:\:\:\:\:\:\:\:\:\:\:&{F_k}\left[ n \right] - \tilde \xi _1^{ - 2}\left[ n \right] \leqslant 0, \forall n,k,&
		\end{align}
		\begin{align}
			\:\:\:\:\:\:\:\:\:&{F_r}\left[ n \right] - \tilde \xi _2^{ - 2}\left[ n \right] \leqslant 0, \forall n,&
		\end{align}
	    \begin{align}
		\:\:\:\:\:\:\:\:\:&\xi _3^{ - 2}\left[ n \right] - {\tilde F_j}\left[ n \right] \leqslant 0, \forall n,&
	    \end{align}	
		\begin{align}
			\:\:\:\:\:\:\:\:\:\:\:\:\:\:&\xi _4^{ - 2}\left[ n \right] - {\tilde F_r}\left[ n \right] \leqslant 0, \forall n,&
		\end{align}	
		\begin{align*}
			\!& (26),(\ref{jammer_power}),  (\ref{GN_power}).&
		\end{align*}	
	\end{spacing}
\end{subequations}
Until now, all constraints are convex, the objective function composed of a linear numerator and a convex denominator. The problem (\ref{Final}) is a quasi-convex optimization problem so that it can thus be solved by employing fractional programming methods, e.g., the Dinkelbach’s algorithm.

\begin{algorithm}[t]
	\renewcommand{\algorithmicrequire}{\textbf{Input:}}
	\renewcommand{\algorithmicensure}{\textbf{Output:}}
	%\caption{Joint trajectory and scheduling optimization algorithm}\label{Algorithm1}
	\caption{An alternating algorithm for solving {(\ref{P0_ee}})}\label{Algorithm1}
	\begin{algorithmic} [1]
		\STATE\textbf{Input:} ${\mu _2}$, ${i_{\max }}$
		\STATE{Initialization:}
		~Set~$i = 0$, as iteration index, ${\mu _2}$ as the threshold and original points ${\Upsilon _0} = \left\{ {{{\mathbf{Q}}^{\left( 0 \right)}},{{\mathbf{\Gamma }}^{\left( 0 \right)}}, {{\mathbf{P}}^{\left( 0 \right)}}, {{\mathbf{U}}^{\left( 0 \right)}},} \right\}$,  thus obtaining the EE by using (\ref{P0_ee}) and generating a series of initial points 
		${\mathbf{S}}_{k,0}^{\left( 0 \right)},{\mathbf{G}}_0^{\left( 0 \right)},{\mathbf{\xi }}_{1,0}^{\left( 0 \right)},{\mathbf{\xi }}_{2,0}^{\left( 0 \right)},{\mathbf{h}}_{ru}^{\left( 0 \right)},{\mathbf{ d}}_{k,0}^{\left( 0 \right)}, {\mathbf{ d}}_{j,0}^{\left( 0 \right)}$.
		\STATE\textbf{~repeat.}
		\STATE~ With given ${{\mathbf{Q}}^{\left( i \right)}}$, ${{\mathbf{\Gamma }}^{\left( i \right)}}$, and ${{\mathbf{U}}^{\left( i \right)}}$, update ${{\mathbf{P}}^{\left( i \right)}}$ to ${{\mathbf{P}}^{\left( i+1 \right)}}$  by solving sub-problem (\ref{P_transmit}).
		\STATE~ With given ${{\mathbf{Q}}^{\left( i \right)}}$, ${{\mathbf{\Gamma }}^{\left( i \right)}}$, update ${{\mathbf{U}}^{\left( i \right)}}$ to ${{\mathbf{U}}^{\left( i+1 \right)}}$  by solving sub-problem (\ref{P_com}).
		\STATE~ With given ${{\mathbf{Q}}^{\left( i \right)}}$, ${{\mathbf{p}}^{\left( {i + 1} \right)}}$, and ${{\mathbf{U}}^{\left( {i + 1} \right)}}$, update ${{\mathbf{\Gamma }}^{\left( i \right)}}$  to ${{\mathbf{\Gamma }}^{\left( i+1 \right)}}$  by solving sub-problem (\ref{P2}). 	
		\STATE~ With given ${{\mathbf{\Gamma }}^{\left( {i + 1} \right)}}$, ${{\mathbf{P}}^{\left( {i + 1} \right)}}$, and ${{\mathbf{U}}^{\left( {i + 1} \right)}}$, update ${{\mathbf{Q}}^{\left( i \right)}}$ to ${{\mathbf{Q}}^{\left( i+1 \right)}}$ by solving sub-problem (\ref{Final}).
		\STATE~ With given ${{\mathbf{Q}}^{\left( i+1\right)}}$, ${{\mathbf{P}}^{\left( i+1 \right)}}$, ${{\mathbf{\Gamma }}^{\left( i+1 \right)}}$, ${{\mathbf{U}}^{\left( {i + 1} \right)}}$, compute EE and ${\mathbf{S}}_0^{\left({i+1}\right)}$, ${\mathbf{T}}_0^{\left( {i + 1} \right)}$, ${\mathbf{\xi }}_{10}^{\left( {i + 1} \right)}$, ${\mathbf{\xi }}_{20}^{\left( {i + 1} \right)}$, ${\mathbf{h}}_{ru}^{\left( {i + 1} \right)}$, ${\mathbf{d}}_{k,0}^{\left( {i + 1} \right)}$, ${\mathbf{d}}_{j,0}^{\left( {i + 1} \right)}$.
		\STATE~ {Update $ i \leftarrow i + 1\ $}.
		\STATE\textbf{~until}: Convergence or $i=i_{\max}$. 	
		\STATE\textbf{Output:}$\left\{ {{{\mathbf{Q}}},{{\mathbf{\Gamma }}}, {{\mathbf{P}}}, {{\mathbf{U}}},} \right\}$. 
	\end{algorithmic}
\end{algorithm}

\subsubsection{Overall Algorithm}
					
In summary, by exploiting the BCD method, we divide the original problem into four blocks. Furthermore, with the aid of Dinkelbach’s algorithm and SCA, four sub-problems are solved as shown. The overall algorithm for solving (\ref{P0_ee}) is summarized in Algorithm 2. 

Note that there are two layers in Algorithm 2, including an inner iteration as shown in Algorithm 1 and an outer iteration. In fact, solving sub-problem 3 and sub-problem 4 dominates the complexity of Algorithm 2. In the inner layer, the fraction programing in sub-problem 3 and 4 can always converge to the stationary and sub-optimal solution \cite{17}. In the outer layer, BCD technique is adopted, and we update $\left\{ {{\bf{P}},{\bf{U}},{\bf{\Gamma }},{\bf{Q}}} \right\}$ in an alternating manner. Consequently, the computational complexities of Algorithm 2 is approximately ${{\cal O}_{sub2}}\left( {N{{ {L }}^{3}}{I_1}{I_2}} \right) + {{\cal O}_{sub3}}\left( {8{N^{3.5}}{I_2}{I_3}} \right)$, where $I_1$ is the number of iterations required for solving (\ref{omiga}),  $I_2$ is for (\ref{Final}), and $I_3$ is for the Dinkelbach's algorithm.

\section{Numerical Results}
	\label{Numerical Results}
	%\vspace{10pt}

\begin{table}
	\centering
	\caption{SIMULATIONS PARAMETERS}
	\begin{tabular}{p{5cm}|p{3cm}}
		\hline\hline
		UAV flight altitulde&$H_u=100$ m\\
		\hline
		Maximum UAV speed&$V_{max}=60$ m/s\\
		\hline	
		Number of IoT devices&$k=5$\\
		\hline
		Element spacing of IRS&$d=0.1$ m\\
		\hline
		Number of IRS elements&$M=150$ \\
		\hline
		Noise power&$\sigma^2=-80$ dBm\\
		\hline
		System bandwidth&$B=1$ MHz\\
		\hline
		Channel power gain at the reference distance&$\rho=-30$ dB\\
		\hline
		Average transmit power&$\bar p=20$ dBm\\
		\hline
		Maximum transmit power&$p_{max}=26$ dBm\\
		\hline
		Jamming power&$\bar p_j=30$ dBm\\
		\hline
		Flight time of UAV&$T=20$ s\\
		\hline
		Time slot&${\Delta t} = 0.5$ s\\
		\hline\hline
	\end{tabular}		
\end{table}

\begin{figure}[t]
	%	\vspace{3pt}	
	\centerline
	{\includegraphics%[width=.36\textwidth]
		[width=1\columnwidth]
		{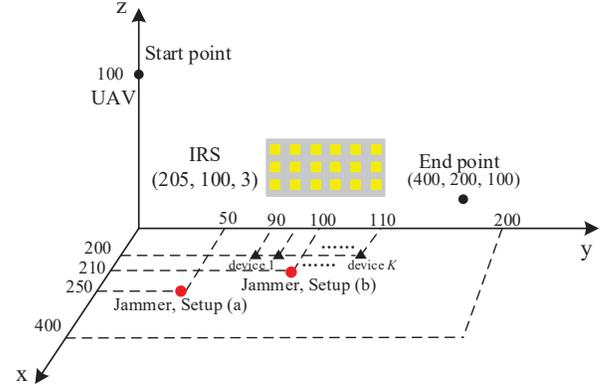}}
	\caption{\label{fig_mod}UAV trajectory for different cases.}
	%	\vspace{4pt}
\end{figure}

In this part, we give the simulation results to verify the effectiveness of the proposed design. {\bf{``Proposed"}} refers to the case we proposed for maximizing the EE by the jointly optimizing communication scheduling,  ID transmit power allocation, IRS passive beamforming, and UAV trajectory. On the other hand, we compared our proposed algorithm with the case without IRS proposed in \cite{anti_jammer}, denoted by {\bf{``w/o IRS"}}. Moreover, we consider two setups for the location of the jammer as shown in Fig. \ref{fig_mod}. For Setup (a), the location of the jammer is ${{\rm{{\bf{q}}}}_j} = \left( {250,50,0} \right)$ m, i.e., it is relatively far away from the cluster of the IDs. While in Setup (b), the location of the jammer is ${{\rm{{\bf{q}}}}_j} = \left( {210,100,0} \right)$ m, i.e., the jammer  is located nearby the cluster of the IDs. The initial and final horizontal coordinates of UAV are set as ${{\rm{{\bf{q}}}}_{S}} = \left( {0,0,100} \right)$ m and ${{\rm{{\bf{q}}}}_{F}} = \left( {400,200,100} \right)$ m, respectively. We assume that line segment from the start point to the end point is used here to initialize the feasible point of the variables. The location of IRS is set as ${{\rm{{\bf{q}}}}_{r}} = \left( {205,100,3} \right)$ m. All the values of required parameters in (\ref{Eu}) are set in \cite{EE_zeng_rot}.

\begin{figure}[t]
	%	\vspace{3pt}	
	\centerline
	{\includegraphics%[width=.36\textwidth]
		[width=1.1\columnwidth]
		{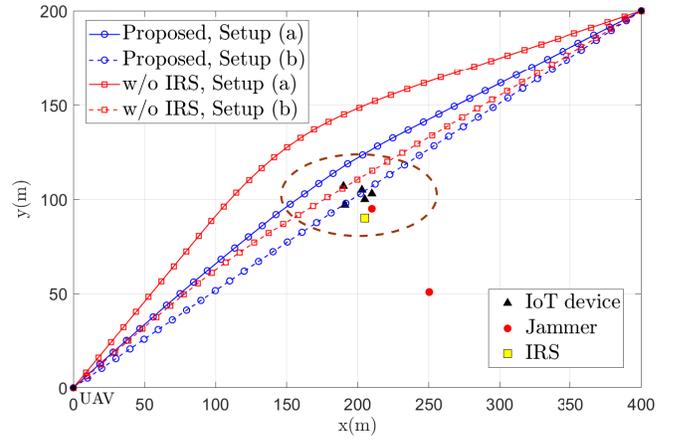}}
	\caption{\label{fig2}UAV trajectories for different cases.}
		\vspace{-10pt}
\end{figure}

\begin{figure*}[htbp]
	%	\vspace*{0pt}
	\subfigure[] {\includegraphics[width=.34\textwidth]{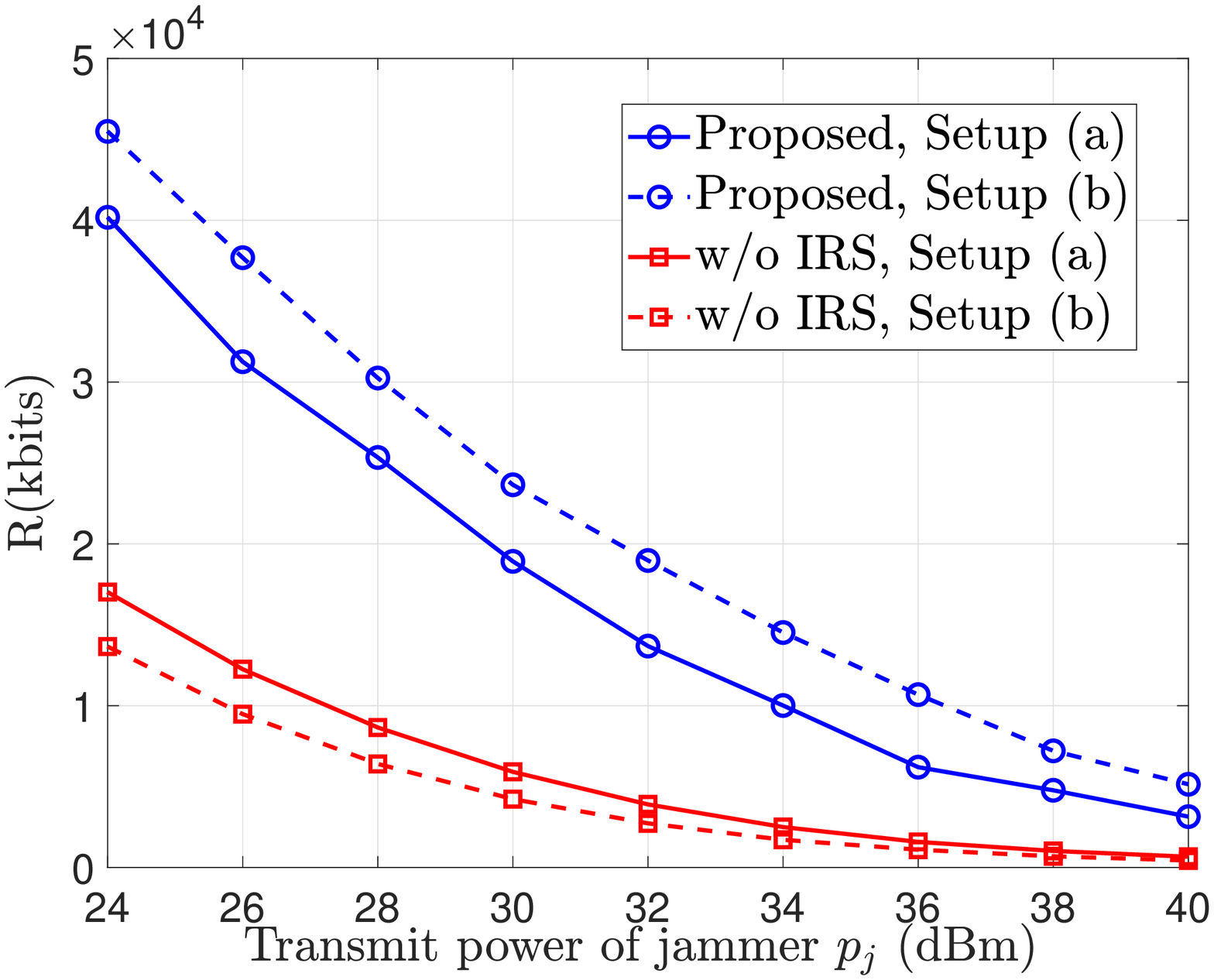}}
	\subfigure[] {\includegraphics[width=.34\textwidth]{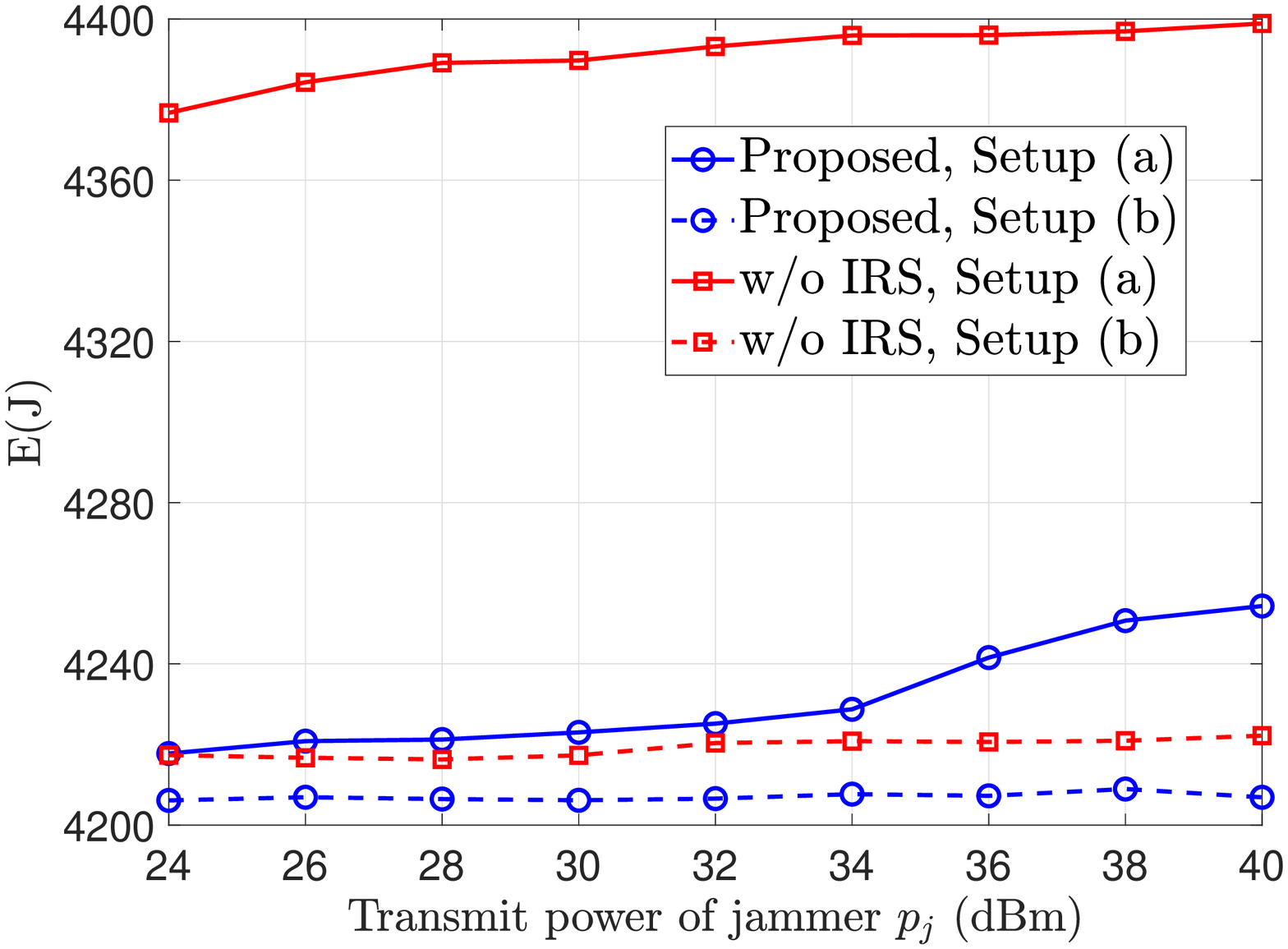}}
	\subfigure[] {\includegraphics[width=.34\textwidth]{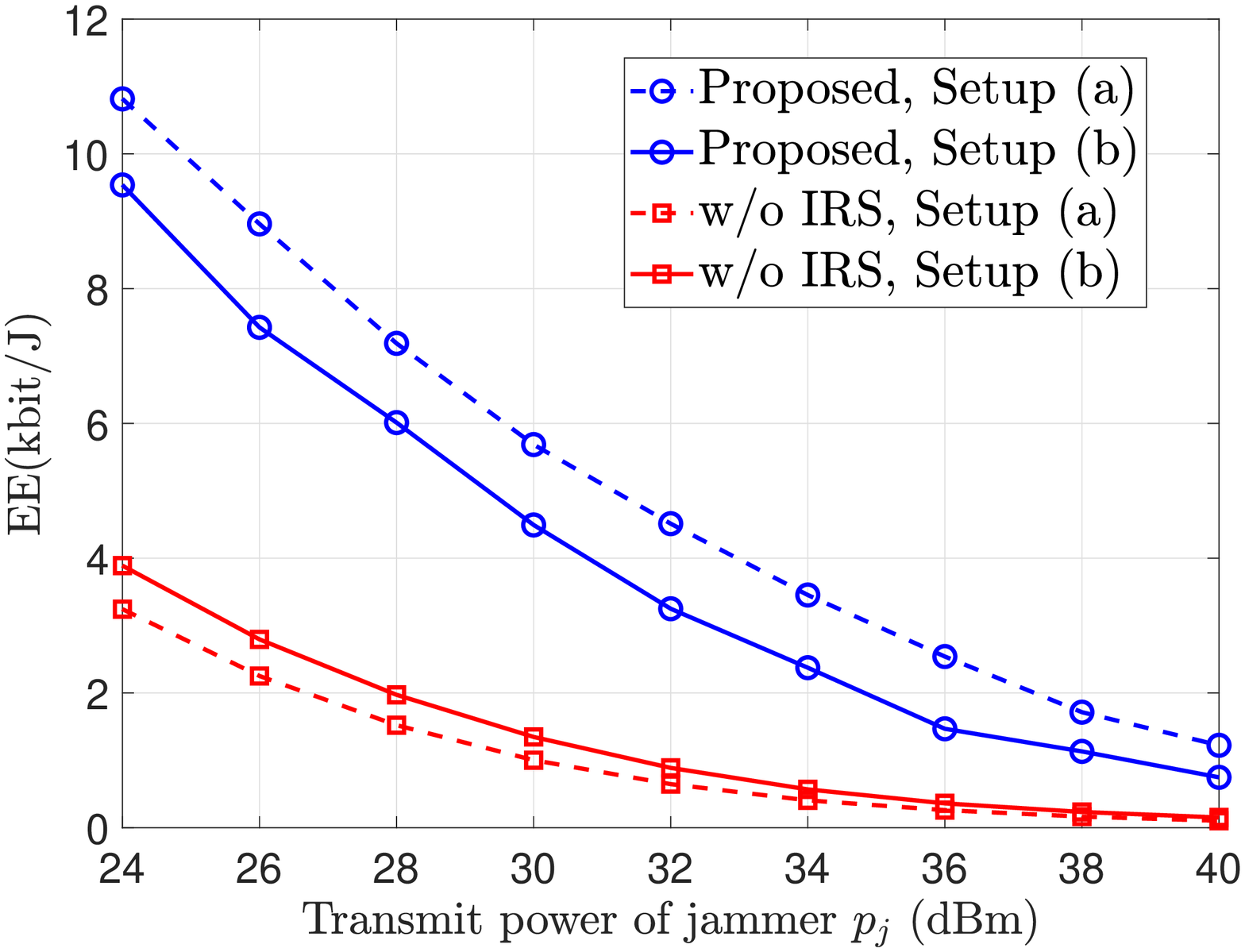}}
	\caption{Performance as (a) the throughput, (b) the energy assumption of UAV, and (c) the EE of the UAV, versus the transmit power of the jammer, i.e., $p_j$, when $M=150$ and $\bar p=20$ dBm.}
	\label{fig3pm}
	%	\vspace*{0pt}
\end{figure*}

Fig. \ref{fig2} shows the UAV's trajectories for different cases. It can be observed that by deploying the IRS to assist in the UAV data collection, our proposed design achieves much shorter flight path than the cases without IRS \cite{anti_jammer}, regardless the location of the jammer. Specifically, in Setup (a), for the case without IRS, the UAV should retreat far away from the jammer for reducing the jamming signal; while for the case with IRS, the received signal at the UAV can be significantly enhanced by the IRS passive beamforming, which can effectively compensate for the jamming and thus enables the UAV to fly much closer to the cluster of IDs. Moreover, one can observe that in Setup (b), the UAV in our proposed design flies straightly from the start point to the end point. The reason is that when the jammer is also located in the coverage of the IRS, the jamming signal can be effectively mitigated at the UAV by IRS passive beamforming, thus the UAV can fly over the IDs for data collection as if there is no jammer. However, for the conventional design without IRS, the retreating away strategy becomes much less effective in Setup (b) (since such behavior would also degrade the signal from the IDs), thus the UAV no more retreats as far as in Setup (a) for saving flight energy.

Fig. \ref{fig3pm} shows the throughput, and energy consumption, and EE versus $p_j$ for different cases. In Fig. \ref{fig3pm}(a), for the benchmark without IRS, Setup (a) facing a remote jammer is more favorable for the UAV data collection than Setup (b) facing a local jammer; however, for the case with IRS, the observation is opposite. The reason is that in our proposed design, when the jammer is located in the coverage of IRS as in Setup (b), the additional gain from  reducing the jamming signal via passive beamforming is achieved except that from enhancing the desired signal as in Setup (a), which thus renders a higher SINR at the UAV. While for the benchmark, the retreating away strategy is effective only in the remote jammer case for reducing the jamming signal; when the jammer is co-located with the IDs, retreating away becomes much less effective because such behavior also degrades the reception of the desired signal. In Fig. 4(b), one can observe that our proposed design achieves lower energy consumption than the benchmark scheme in both setups. Moreover, in Setup (b), the energy consumption for both our proposed and the benchmark designs almost keeps unchanged as $p_j$ increases. This is because in our proposed design, the UAV flies straightly from the start point to the end point regardless of $p_j$ (thanks to the effectively jamming mitigation); while in the benchmark design, the UAV also would not retreating further as $p_j$ increases since the retreating away strategy becomes less effective. On contrast, in Setup (a) facing remote jammer, retreating away is beneficial for increasing the SINR for both designs. Owing to the above, we can conclude that by deploying the IRS nearby the IDs for assisting in the UAV data collection, the EE of the UAV can be significantly improved, especially for the case when the jammer is co-located with the IDs (deemed challenging in the conventional systems without IRS), which is also validated by the observation in Fig. 4(c).  
		
\begin{figure}[t]
	%	\vspace{3pt}	
	\centerline
	{\includegraphics%[width=.36\textwidth]
		[width=1.1\columnwidth]
		{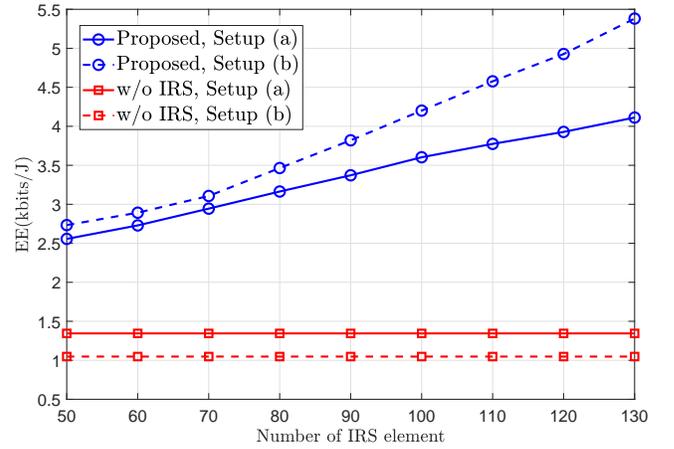}}
	\caption{\label{fig4M}The EE of UAV versus the number of IRS elements, i.e., $M$.}
%		\vspace{-10pt}
\end{figure}

In Fig. \ref{fig4M}, the EE of the UAV for different cases versus the number of IRS elements is illustrated. It is observed that with the increasing of IRS element $M$, the EE for the cases with IRS are both improved, which validates the performance gain by enlarging the IRS size. Moreover, one can observe that in our proposed design, Setup (b) achieves higher EE of UAV than Seutp (a) (the reason is as similar as that in Fig. 4 and thus omitted here) and the performance gap enlarges as $M$ increases. This is because a larger number of IRS elements enables a larger jamming signal mitigation gain in Setup (b), which is however unavailable in Setup (a).

%\begin{figure}[t]
%	%	\vspace{3pt}	
%	\centerline
%	{\includegraphics%[width=.36\textwidth]
%		[width=0.95\columnwidth]
%		{fig5dm1.eps}}
%	\caption{\label{fig6dm} EE for different cases versus the distance between jammer and cluster.}
%	%	\vspace{4pt}
%\end{figure}

\begin{figure}[t]
	%	\vspace{3pt}	
	\centerline
	{\includegraphics%[width=.36\textwidth]
		[width=1.1\columnwidth]
		{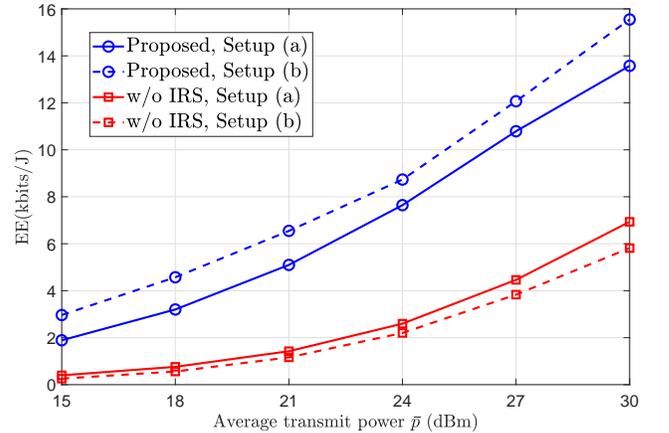}}
	\caption{\label{fig7pop}The EE of UAV versus the average transmit power of IDs, i.e., $\bar p$ .}
%		\vspace{-10pt}
\end{figure}

Fig. \ref{fig7pop} shows the EE of the UAV versus the average transmit power of IDs $\bar p$. It can be observed that as $\bar p$ increases, the EE of UAV for all cases are improved since the received signal at the UAV are enhanced by increasing the uplink transmit power. More importantly, one can observe that in Setup (a), for achieving the same EE of UAV as 4 kbits/J, a transmit power of $26$ dBm is required at the IDs for the benchmark while this value is reduced to $19$ dBm for our proposed design, which suggest a 7 dB gain by deploying the IRS nearby the IDs. Note that such performance improvement even reaches up to almost 10 dB in Setup (b) due to the additional jamming mitigation gain. The above observation reveals that the proposed design can not only help to improve the EE of UAV, but also is beneficial for saving the transmit power of the IDs and thus prolong their life time.

%\vspace{5pt}
\section{Conclusions}\label{Conclusions}
%\vspace{3pt}
					
In this paper, we study an IRS-aided UAV data collection system in the presence of a jammer. By considering the communication scheduling, IDs' power allocation, IRS passive beamforming, and UAV trajectory, an alternating optimization based algorithm is proposed to solve the problem by exploiting the SCA and BCD techniques. Simulation results showed that by deploying the IRS in our proposed design, the EE of the UAV has improved significantly, especially for the case facing a local jammer, which is deemed highly challenging in the conventional systems without IRS.

\bibliographystyle{IEEEtran}
					%\balance
%\vspace{-5pt}				

\end{document}